\documentclass[prl,aps,twocolumn,showpacs,floatfix]{revtex4-1}
\usepackage{amsfonts}
\usepackage{amssymb}
\usepackage{hyperref}
\usepackage{graphicx}
\usepackage{dcolumn}
\usepackage{bm,amsmath,verbatim}
\usepackage{mathrsfs}
\usepackage{color}

\hypersetup{colorlinks,
linkcolor=blue,          citecolor=blue,        filecolor=blue,      urlcolor=blue           }

\def\be{\begin{equation}}
\def\ee{\end{equation}}
\def\bea{\begin{eqnarray}}
\def\eea{\end{eqnarray}}
\def\bse{\begin{subequations}}
\def\ese{\end{subequations}}

\def\be{\begin{eqnarray}}
\def\ee{\end{eqnarray}}

\begin{document}

\title{Dynamical Weyl Points and 4D Nodal Rings in Cold Atomic Gases}
\author{Yan-Bin Yang$^1$}
\author{L.-M. Duan$^{1,2}$}
\author{Yong Xu$^{1}$}
\email{yongxuphy@tsinghua.edu.cn}
\affiliation{$^1$ Center for Quantum Information, IIIS, Tsinghua University, Beijing 100084, PR China}
\affiliation{$^2$ Department of Physics, University of Michigan, Ann Arbor,
Michigan 48109, USA}

\begin{abstract}
Controllability of ultracold atomic gases has reached an unprecedented level,
allowing for experimental realization of the long-sought-after Thouless pump, which
can be interpreted as a dynamical quantum Hall effect. On the other hand,
Weyl semimetals and Weyl nodal line semimetals with touching points and rings in
band structures have sparked tremendous interest in various fields in the past few years.
Here, we show that dynamical Weyl points and dynamical 4D Weyl nodal rings, which
are protected by the first Chern number on a parameter
surface formed by quasi-momentum and time,
emerge in a two-dimensional and three-dimensional system, respectively. We find that the topological
pump occurs in these systems but the amount of pumped particles is not quantized and
can be continuously tuned by controlling experimental parameters over a wide range.
We also propose an experimental scheme to realize
the dynamical Weyl points and 4D Weyl nodal rings and to observe their corresponding topological pump in cold atomic gases.
\end{abstract}
\maketitle

Recently, topological gapless phenomena have seen a rapid advance
in three-dimensional (3D) condensed matter systems ranging from solid-state materials~\cite{Wan2011prb, Yuanming2011PRB, Burkov2011PRL, ZhongFang2011prl,Shengyuan2014PRL,Soluyanov2015Nature,Ueda2016,LiangFu2016}, cold
atoms~\cite{Gong2011prl, Anderson2012PRL, Yong2014PRL, Tena2015RPL, Bo2015,Yong2015PRL, Yong2016typeii} to optical and acoustic systems~\cite{LingLu2013NP,Meng2015}. This is mainly attributed to their powerful ability
to simulate fundamental physics~\cite{Burkov2016, Jia2016, VishwanathRMP}. For instance, Weyl fermions,
which have been long-sought-after in particle physics, have recently been experimentally
observed in condensed matter systems~\cite{Lv2015,Xu2015,Lu2015,Shuyun2016,Huang2016}.
These fermions protected by the first
Chern number can be viewed as the quantum Hall phase transition points in the momentum space,
leading to an anomalous Hall effect~\cite{VishwanathRMP}. Another celebrated example of 3D
gapless phenomena is the Weyl nodal ring~\cite{Burkov2011PRB,HasanRing,Yong2016PRA,DWZhang2016}, which has also been experimentally observed recently~\cite{HasanRing}.
Even though they are topologically protected by the quantized Berry phase, the
anomalous Hall effect cannot occur in this system in the absence of external magnetic fields.

On the other hand, Thouless predicted the quantized transport
of particles arising from a cyclic deformation of an underlying Hamiltonian without an
applied bias voltage~\cite{Thouless1983}, which has recently been observed in cold
atom experiments~\cite{Nakajima2016,Bloch2016}, thanks to rapid progress of cold atom
technology. Such a quantized transport can be interpreted
as the dynamical quantum Hall effect~\cite{PolkovnikovPNAS,Bloch2016}
on a surface formed by quasi-momentum and time. Given that the anomalous Hall effect occurs in
Weyl semimetals, a natural question to ask is whether a dynamical Weyl point featuring
a topological transport that can be interpreted as a dynamical anomalous Hall effect
exists. Since Weyl nodal semimetals in 3D do not support the anomalous Hall effect, we do not
expect the existence of a dynamical Weyl nodal ring. However, in a 3D system, viewing time as a
parameter, one may wonder whether a new dynamical gapless phenomenon featuring the topological transport
appears.

In this paper, we demonstrate that the dynamical Weyl points can be engineered in a 2D
slowly periodically-driven system. Here, besides two quasi-momenta, e.g., $k_x$ and $k_y$,
time $t$ may be regarded as an artificial parameter, taking the place of another quasi-momentum
parameter $k_z$. When the adiabatic condition is fulfilled, the Weyl point can be
characterized by the Chern number defined on a closed surface in the space $(k_x,k_y,t)$
enclosing the point. Alternatively, because of the periodicity of the system,
it ends up with the same state over a cycle, implying that the system at time $t$
is equivalent to that at time $t+T$ with $T$ being the period, reminiscent of the property
of a Brillouin zone. Hence, the Chern number can be defined on a torus $(k_x,t)$
like in the momentum space. In a 3D system, adding the time as a parameter
gives us a 4D system and a dynamical 4D Weyl nodal ring emerges.
Different from a Weyl nodal ring~\cite{Burkov2011PRB,Yong2016PRA,DWZhang2016} in
the 3D space that is protected by a quantized Berry phase,
the 4D Weyl nodal ring is characterized by the first Chern number.

\begin{figure}[t]
\includegraphics[width=3.2in]{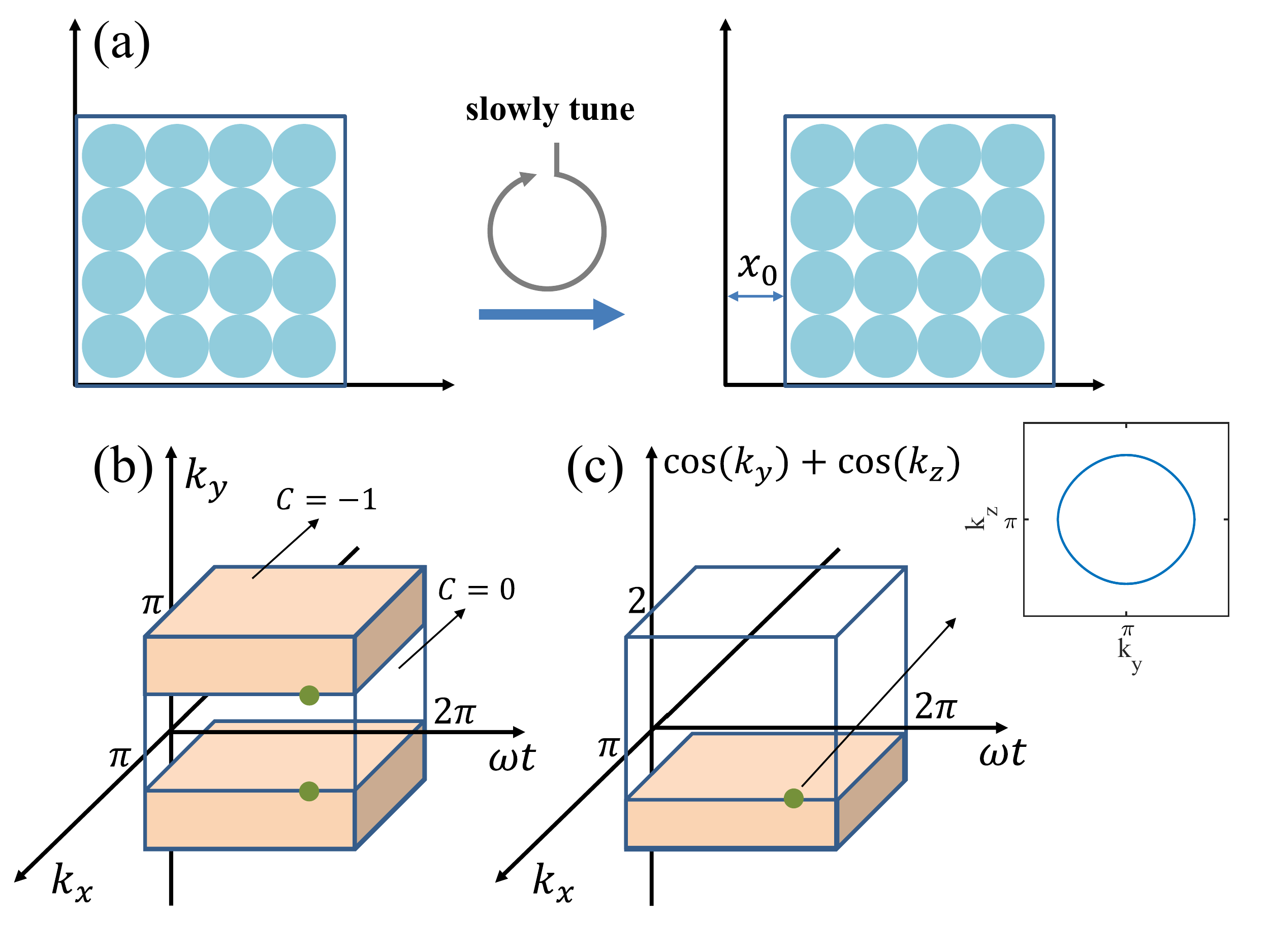}
\caption{\textbf{Schematic of 2D pumping and locations of dynamical Weyl points
and 4D Weyl nodal rings.} \textbf{(a)} Sketch of pumping particles in 2D by slowly varying
parameters of a system without a bias, where the average displacement of a cloud of atoms
over an entire cycle is denoted by $x_0$. \textbf{(b)} and \textbf{(c)} Distribution of the
Chern number defined in the $(k_x,\omega t)$ plane
as a function of $k_y$ in 2D and $\cos(k_y)+\cos(k_z)$ in 3D, respectively. The Chern number
is $-1$ in the light orange regions and $0$ in other regions.
The green points in (b) and (c) represent the dynamical Weyl points and 4D Weyl nodal rings [shown in
the inset in (c)], respectively; they separate the topological trivial and nontrivial phases.
In (b), $\lambda=1$ and $M_0=2$;
In (c), $\lambda=1$ and $M_0=3$.}
\label{Fig1}
\end{figure}

Furthermore, we show that the dynamical Weyl points and the 4D Weyl nodal rings give rise to
a non-quantized topological pump [as schematically illustrated in Fig.~\ref{Fig1}(a)] and the amount of pumped particles can be continuously tuned
by controlling the experimental parameters, similar to the classical Archimedes screw that can be
tuned by tilting the screw, even though the physics underlying our system is quantum mechanics
and topology. Finally, we propose an experimental scheme to realize the dynamical Weyl points
and 4D Weyl nodal rings and to observe their corresponding tunable topological
pump in cold atomic gases.

{\LARGE \textbf{Results}}

\textbf{Model Hamiltonian.} Consider a toy model that is described by the following time-dependent Hamiltonian
in the momentum space
\begin{eqnarray}
H(t)=&&-\sin(k_x)\sigma_x+\lambda\cos(\omega t)\sigma_y+[M+\cos(k_x)+ \nonumber \\
&&\lambda\sin(\omega t)]\sigma_z,
\label{toymodel}
\end{eqnarray}
where $k_x$ is the quasi-momentum in the x direction, $\sigma_\nu$ with $\nu=x,y,z$ are the Pauli matrices,
and $\lambda$ is a real parameter (we take $\lambda>0$ for simplicity).
Here, $M=M_0+\cos(k_y)$ with $M_0$ being a real parameter in 2D and
$M=M_0+\cos(k_y)+\cos(k_z)$ in 3D, where $k_y$ and $k_z$ are the quasi-momenta in the $y$ and $z$
directions, respectively. The unit of energy and length is taken to be 1. The Hamiltonian is
time-dependent and periodic with $H(t+T)=H(t)$ and $T=2\pi/\omega$.

Provided $M=M_0$, this Hamiltonian is a typical model of a Chern band~\cite{Xiaoliang2008}
in the ($k_x,k_y$) space if $\omega t$ is replaced with $k_y$. Instead, we define
the Chern number in the ($k_x,t$) space for the $n$th instantaneous band as
\begin{equation}
C_n(M)=\frac{1}{2\pi}\int_{-\pi}^{\pi} dk_x \int_{0}^{T} dt \Omega_n (k_x,t),
\end{equation}
where $t$ takes the place of a quasi-momentum, and the Berry curvature~\cite{XiaoRMP} is
$\Omega_n(k_x,t)=-2\text{Imag}(\langle\partial_{k_x} u_n(k_x,t)|\partial_t u_n(k_x,t)\rangle)$ with
$|u_n(k_x,t)\rangle$ being the $n$th instantaneous eigenstate of $H(t)$, i.e., $H(t)|u_n(k_x,t)\rangle=E_n(k_x,t)|u_n(k_x,t)\rangle$.

By straightforward calculation, we find $C_1=1$ if $-(1+\lambda)<M<-|1-\lambda|$,
$C_1=-1$ if $|1-\lambda|<M<1+\lambda$, and zero, otherwise.
In 2D, the Chern number changes abruptly with respect to $k_y$,
implying a transition between different dynamical quantum Hall phases
in the momentum space. The transition point may therefore be called the dynamical Weyl point.
These points are located at $[k_x^W=0,k_{y}^W=\alpha \text{arccos}(-1-M_0\mp\lambda),\omega t=\pm \pi/2]$ with $\alpha=\pm1$ when $-2\mp\lambda<M_0<\mp\lambda$, and at $[k_x^W=\pi,k_y^W=\alpha \text{arccos}(1-M_0\mp\lambda),\omega t=\pm \pi/2]$
when $\mp\lambda<M_0<2\mp\lambda$. For instance,
when $\lambda=1$ and $M_0=2$, with the second condition being satisfied, there
appear two gapless points located at $(k_x^W=\pi,k_y^W=\pm \pi/2,\omega t=3\pi/2)$,
as shown in Fig.~\ref{Fig1}(b). These points correspond to the abrupt change of the Chern number
along $k_y$, i.e., $C_1=-1$ when $|k_y|>\pi/2$ and $C_1=0$, otherwise, as displayed in Fig.~\ref{Fig1}(b).
Alternatively, one may choose a closed surface enclosing the point
and find its Chern number equal to $\pm 1$.

In 3D, we have a 4D space characterized by $(k_x,k_y,k_z,t)$, if viewing $t$ as a parameter.
We find gapless rings lying
in the $(k_x^W=0,\omega t=\pm \pi/2)$ plane when $-3\mp\lambda<M_0<1\mp\lambda$ or in the
$(k_x^W=\pi,\omega t=\pm \pi/2)$ plane when $-1\mp\lambda<M_0<3\mp\lambda$.
For example, when $\lambda=1$ and $M_0=3$, a single gapless ring appears in the $(k_y,k_z)$
plane corresponding to $k_x=\pi$, $\omega t=3\pi/2$ and $\cos(k_y)+\cos(k_z)=-1$, as illustrated in
Fig.~\ref{Fig1}(c). In contrast to a Weyl nodal ring that is protected by the quantized Berry phase~\cite{Burkov2011PRB,Yong2016PRA,DWZhang2016},
this ring is characterized by the first Chern number over a closed surface enclosing the ring.
We therefore dub it a dynamical 4D Weyl nodal ring. In Fig.~\ref{Fig1}(c), we also show that the ring
corresponds to the topological phase transition of dynamical quantum Hall effects.
Inside a ring for a fixed $(k_y, k_z)$, the Chern number over the $(k_x,\omega t)$ torus
is $-1$; outside the ring, it is 0.

\textbf{Topological pump.} With the dynamical Weyl points and 4D Weyl nodal rings, we are now ready to
study the topological pump in these systems.
The number of pumped particles per unit
length in 2D or per unit area in 3D is given by~\cite{XiaoRMP}
\begin{equation}
N_p=\sum_n \int_{0}^T dt\int_{BZ} \frac{d{\bf k}}{(2\pi)^{d}}\langle \psi_n({\bf k,t})|\hat{{\bf v}}|\psi_n({\bf k,t})\rangle,
\end{equation}
where $\hat{{\bf v}}=\partial_{\bf k}H({\bf k})$ is the velocity operator, $d$ is the
dimension of a system, and $|\psi_n({\bf k,t})\rangle$ is the evolution of a state initialized to the $n$th
eigenstate of $H(0)$; the integral in the momentum
space is over a Brillouin zone and $\sum_n$ is the summation over the filled bands.

\begin{figure}[t]
\includegraphics[width=3.4in]{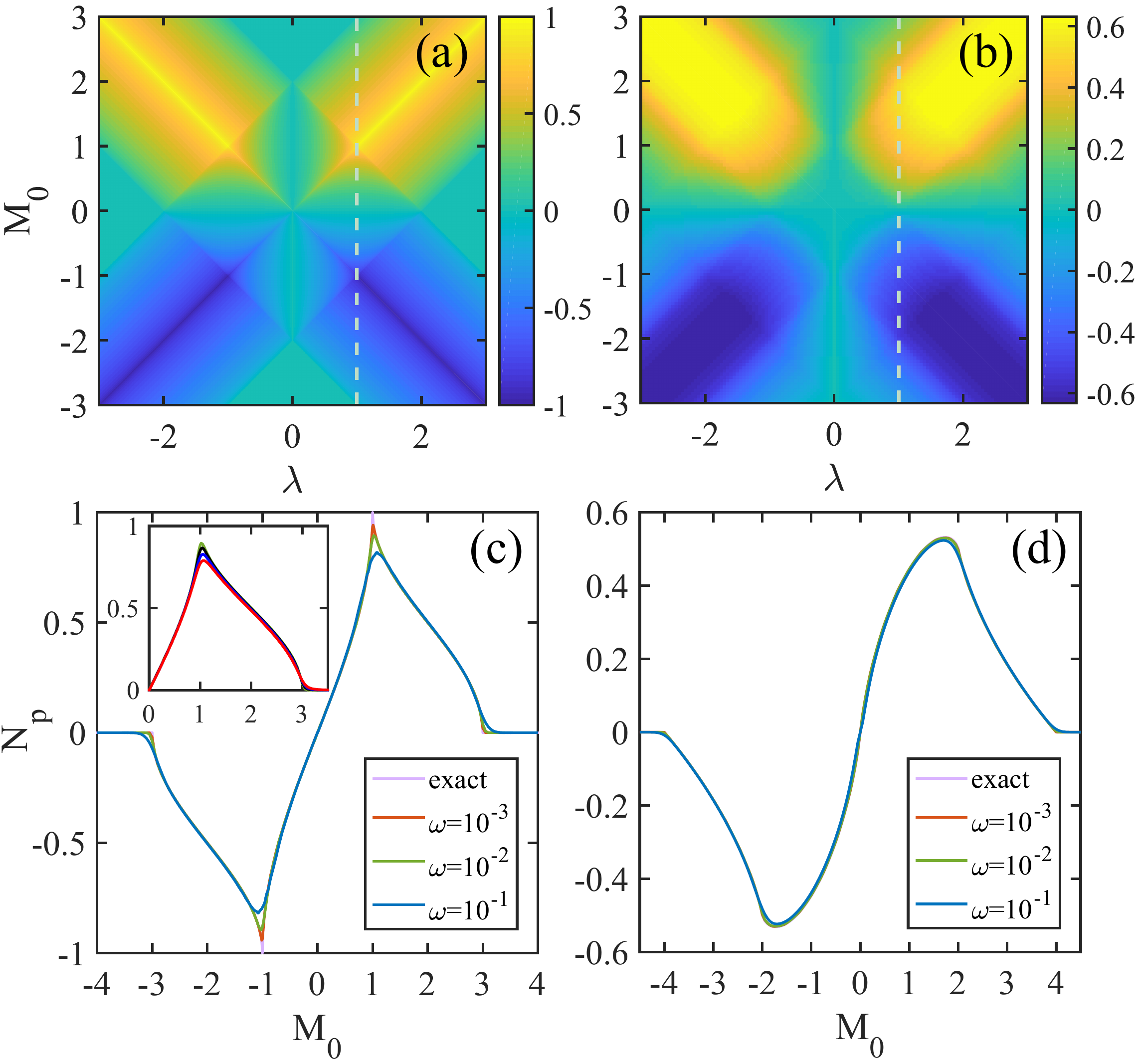}
\caption{ \textbf{Amount of pumped particles.} Amount of pumped particles \textbf{(a)} per unit length in 2D
and \textbf{(b)} per unit area in 3D with respect to $\lambda$ and $M_0$. The results are
obtained under the adiabatic condition. Amount of pumped particles \textbf{(c)} per unit
length in 2D and \textbf{(d)} per unit area in 3D for $\lambda=1$, which are numerically calculated
for $\omega=0.001, 0.01, 0.1$. The exact result corresponds to the case with $\omega\rightarrow 0$.
The inset in (c) plots the results for $\omega=0.01$ and the dephasing rate $\gamma=0,0.05,0.1,0.15$
as the green, black, blue and red lines, respectively. }
\label{Fig2}
\end{figure}

With the assumption of the adiabatic condition (i.e., $\omega$ is sufficiently small),
by the time-dependent perturbation theory, the formula above can be reduced to
\begin{equation}
N_p=-\sum_n \int_{BZ^\prime} \frac{d{\bf k}^\prime}{(2\pi)^{d-1}}C_{n}({\bf k}^\prime),
\end{equation}
where the integration is performed over the momentum space except $k_x$. In 1D,
it corresponds to the celebrated result obtained by Thouless~\cite{Thouless1983}.
In higher dimensions, the formula indicates that the amount of pumped particles
is dictated by the length in 2D (area in 3D) with corresponding Chern numbers.
As a consequence, the amount is not necessarily quantized and its value can be
tuned by changing the length in 2D and area in 3D. In our toy model, it
is determined by $M_0$ and $\lambda$. For example, in 2D,
the amount depends on the distance between two dynamical Weyl points along $k_y$.

To demonstrate how the pump can be tuned, we plot the amount of pumped particles
per unit length in 2D and per unit area in 3D over a cycle with respect to
$M_0$ and $\lambda$ in Fig.~\ref{Fig2}. In the ideal case with an infinitesimal
$\omega$, the amount can be tuned from -1 to 1 in 2D and from -0.63 to 0.63 in 3D.
It is symmetric and antisymmetric with respect to $\lambda=0$ and $M_0=0$, respectively;
the antisymmetry reflects the flip of the charge of the dynamical Weyl points and 4D Weyl nodal rings.
Because of the presence of these gapless points (or rings), one may wonder whether
the excitation near the gapless regions would compromise our results. To check this,
we perform the numerical calculation of the amount of pumped particles using distinct
finite $\omega$ for $\lambda=1$ and plot the results in Fig.~\ref{Fig2}(c) and (d).
They show that the influence on the transport over a cycle is very small in most parts
except in the vicinity of $M_0=\pm 1,\pm 3$, where $N_p=\pm1, 0$ in 2D, and in almost the
whole region in 3D, even when $\omega=0.1$. The nonadiabaticity effects are directly
related to the probability that particles are excited to the higher band near the
gap closing region. Around these regions, the Hamiltonian is approximated by
$H=-\sin(k_x)\sigma_x+(M\pm1+\cos(k_x))\sigma_z\mp\omega\delta t\sigma_y$,
where $\delta t$ is measured with respect to $t=\pm\pi/(2\omega)$.
According to the Landau-Zener formula,
the total number of excited particles into a higher band is given by $N_e\approx \int_0^\infty dE D(E)P(E)$,
where $D(E)$ is the density of states and
$P(E)=e^{-\pi E^2/\omega}$.
In 2D, when
$M_0=1$ or $M_0=3$, with $E\approx \sqrt{\delta k_x^2+(\delta k_x^2\pm\delta k_y^2)^2/4}$
near the gapless points, we can qualitatively assume $E\approx \sqrt{\delta k_x^2+\delta k_y^4/4}$
(which is fulfilled when $\delta k_y \gg \delta k_x$) and find $D(E)\propto \sqrt{E}$. Yet,
in other regions, $E\approx\sqrt{\delta k_x^2+
\sin(k_{y}^W)^2\delta k_y^2}$ and $D(E)\propto E$. Apparently,
the number of excited particles in the former case is larger than that in the
latter near zero energy because of higher density of states, leading to the manifest
nonadiabaticity effects. For a dynamical 4D Weyl nodal ring in a 3D system, the
nonadiabaticity effects are also small since $D(E)\propto E$.

\textbf{Dephasing effect.} In a realistic cold atom experiment, a dephasing may appear naturally due to
laser noise. To see whether the topological pump is stable against the dephasing,
which randomizes the coherent superposition of excited and ground states,
we solve a minimal pure-dephasing model described by
the following master equation in the Lindblad form~\cite{Breuer2007}
\begin{equation}
\dot{\rho_{\bf k}}=-i[H(t),\rho_{\bf k}]+\gamma(\bar{\sigma}_z(t)\rho_{\bf k}\bar{\sigma}_z(t)-\rho_{\bf k}),
\end{equation}
where $\rho_{\bf k}$ is the density matrix, $\gamma$ is the dephasing rate,
$\bar{\sigma}_z(t)={\bf d}(t)\cdot{\bm \sigma}/d(t)$ if the Hamiltonian is written as
$H={\bf d}(t)\cdot {\bm \sigma}$. Here, we have adopted a simplest pure-dephasing model
where the Lindblad operator $\bar{\sigma}_z(t)$ is assumed to always commute with the
Hamiltonian $H(t)$. In the inset of Fig.~\ref{Fig2}(c), we plot the amount of pumped particles
per unit length over a cycle in the 2D case as a function of $M_0$ for $\lambda=1$.
The transport is only slightly reduced for small dephasing rates in most parts
and this reduction increases with $\gamma$ as dephasing decreases the transported
amount in each 1D insulator with a fixed $k_y$~\cite{Jiangbin2015}. The reduction
is especially manifest around $M_0=1$, where more particles near the
gapless point are excited to the higher band; these particles lose their coherence by dephasing
and lead to strong suppression of the transport.

\begin{figure}[t]
\includegraphics[width=3.4in]{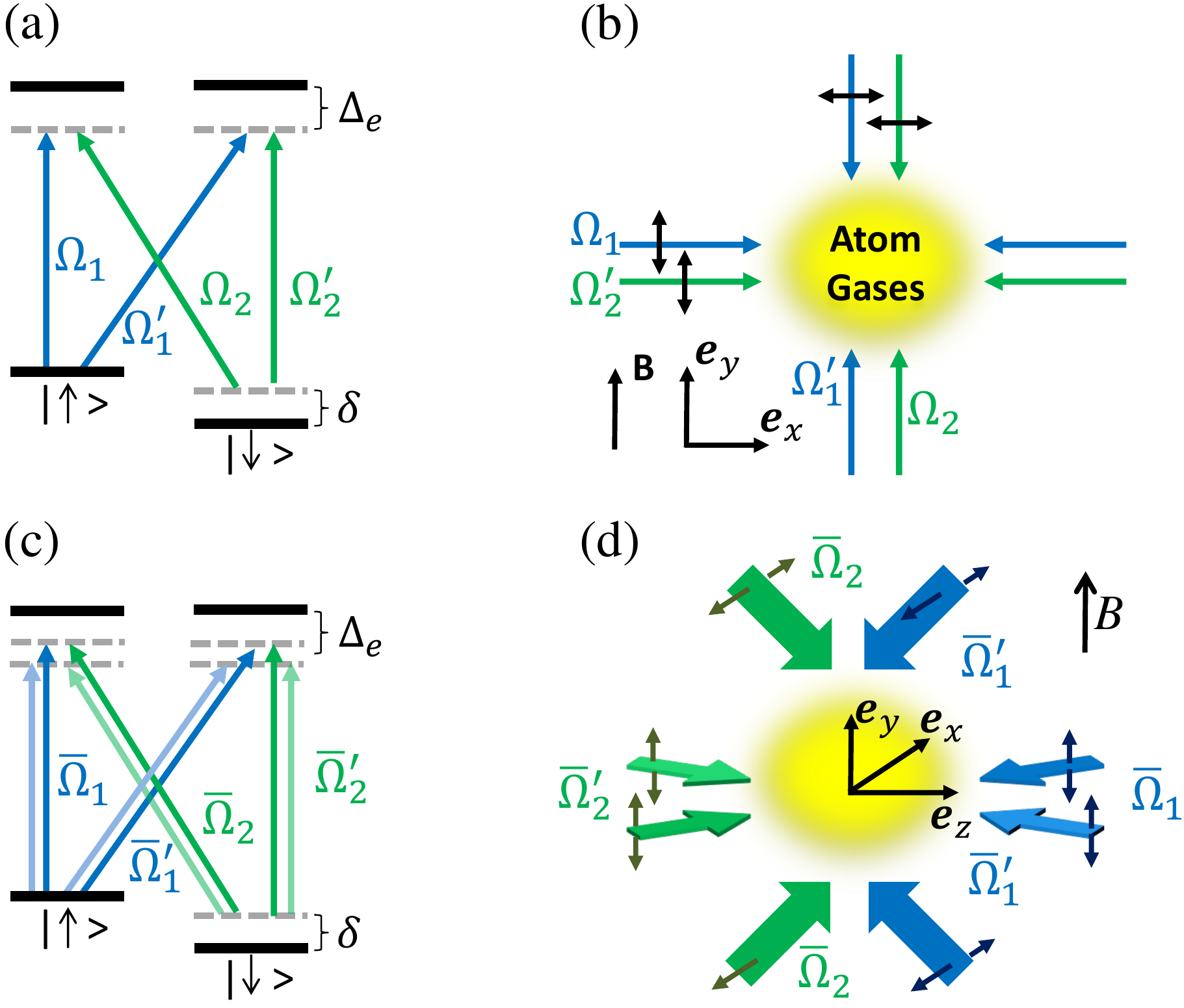}
\caption{\textbf{Laser configurations.} Sketch of laser configurations to realize
\textbf{(a)} and \textbf{(b)} dynamical Weyl points and \textbf{(c)} and \textbf{(d)} dynamical 4D Weyl nodal rings.
The laser beams denoted by the same color arrows
possess the same frequency. The laser beam denoted by the green arrow is generated by applying an acoustic-optical modulator
(AOM) to the other laser beam.
The double arrows describe the linear polarization direction of laser beams.
$\bm B$ is the magnetic field and $\delta$ is the double photon detuning. In (c),
the laser beams with the light colors correspond to those with Rabi frequencies
$\tilde{\Omega}_n$ and $\tilde{\Omega}_n^\prime$ with $n=1,2$. In (d), the configuration
of these lasers is the same as those plotted and thus neglected for clarity.
}
\label{Fig3}
\end{figure}

\textbf{Experimental realization.} To realize the dynamical Weyl points and 4D Weyl nodal rings and their
corresponding tunable topological pump,
we consider the following continuous model
\begin{equation}
H_{C}=\frac{{\bf p}^{2}}{2m}-\sum_{\nu}V_{\nu}\cos^{2}(k_{L\nu}r_{\nu})+h_{z}\sigma_{z}+V_{N}\sigma_{y},
\label{HC}
\end{equation}
where $m$ is the mass of atoms, ${\bf p}=-i\hbar\nabla$ is the momentum operator,
$h_z$ is the Zeeman field, $V_{\nu}>0$ with $\nu=x,y$ in 2D ($\nu=x,y,z$ in 3D) denote the strength of optical lattices with the lattice constants
being $a_\nu=\pi/k_{L\nu}$, and $V_N=V_{SO}+V_{Zy}$ with $V_{SO}=\Omega_{SO}\sin(k_{Lx}r_{x})\cos(k_{Ly}r_{y})$
and $V_{Zy}=-V_{Zy0}\cos(k_{Ly}r_{y})\cos(k_{Lx}r_{x})$ in 2D [
$V_{SO}=\Omega_{SO}\sin(k_{Lx}r_{x})\cos(k_{Ly}r_{y})\cos(k_{Lz}r_{z})$
and $V_{Zy}=-V_{Zy0}\cos(k_{Ly}r_{y})\cos(k_{Lx}r_{x})\cos(k_{Lz}r_{z})$ in 3D].
This model gives us the tight-binding Hamiltonian in the momentum space (see
Methods for details)
\begin{equation}
H({\bf k})=(h_{z}+h_t)\sigma_{z}
-2J_{SO}\sin(k_{x}a_{x})\sigma_{x}+h_{y}\sigma_{y},
\end{equation}
where $h_t=2\sum_{\nu}J_{\nu}\cos(k_{\nu}a_{\nu})$ with $\nu$ being summed over $x$ and $y$ in 2D ($x$, $y$ and $z$ in 3D).
By slowly driving the Hamiltonian according to
$h_y=\lambda\cos(\omega t)$ and $h_z=M_0+\lambda \sin(\omega t)$, we achieve the
toy model in Eq.~(\ref{toymodel}).

To engineer the continuous model in Eq.~(\ref{HC}) in experiments, we can apply the current experimental
technology that implements the 2D spin-orbit coupling in cold atomic gases~\cite{Shuai2015,Shuai2017,Xiongjun2014PRL},
where the spin is represented by two hyperfine states of alkali atoms such as
$^{40}$K~\cite{Jing2015Nature,Jing2015} and $^{87}$Rb~\cite{Shuai2015,Shuai2017}. In Fig.~\ref{Fig3}(a) and (b),
we plot a schematic of a simple and feasible laser configuration scheme for realization of the dynamical Weyl points
exhibiting the tunable topological pumping in a 2D system. Here, two independent sets of linearly polarized Raman laser beams that
couple two states are applied to create the off-diagonal spin-dependent optical lattices.
These lasers have the Rabi frequencies:
$[\Omega _{1}=\Omega _{10}\sin(k_{R}r_{x}),\Omega _{2}=-i\Omega _{20}\cos (k_{R}r_{y})]$
with $k_R$ being the wavevector of the lasers
and $[\Omega _{1}^{\prime }=\Omega _{10}^\prime\cos(k_{R}r_{y}),\Omega _{2}^{\prime }
=i\Omega _{20}^\prime\cos(k_{R}r_{x})]$, respectively, yielding
$\Omega _{SO}=\Omega_{10}^*\Omega _{20}/\Delta _{e}$ and $V_{Zy0}=
\Omega_{10}^{\prime *}\Omega _{20}^\prime/\Delta _{e}$ through Raman processes.
In addition, due to the stark effects, these laser beams generate the spin-independent optical lattices: $%
-V_{x}\cos^{2}(k_R r_x)$ and $-V_{y}\cos^{2}(k_{R}r_{y})$ with $%
V_{x}=(|\Omega _{20}^\prime|^{2}-|\Omega _{10}|^{2})/|\Delta _{e}|$
and $V_y=(|\Omega_{20}|^2+|\Omega_{10}^\prime|^2)/|\Delta_e|$.
In an experiment, one may choose $^{40}$K atoms and use a red-detuned laser beam with
wavelength $773$nm~\cite{Jing2015Nature}, yielding the recoil energy
$E_R=\hbar^2 k_R^2/2m=2\pi\times 8.3\text{kHz}$.
Taking
$|\Omega_{20}|=|\Omega_{20}^\prime|=2\pi\times 0.244 \text{GHz}$ and
$|\Omega_{10}|=|\Omega_{20}^\prime|/3$, we have $V_x=4.9E_R$,
$J_x=0.08E_R$ and $h_y=-0.72V_{Zy0}$.
To implement the pump, we should vary $h_y$ and $h_z$ according to
$h_y=2J_x\cos(\omega t)$ and $h_z=M_0+2J_y \sin(\omega t)$ by controlling the strength of the lasers
and plugging a $\pi$ phase appropriately by a AOM, and by controlling the frequency of the
lasers represented by the green arrows in Fig.~\ref{Fig1} as $h_{z}=\delta /2$, respectively.
Note that when $h_y=2J_x$, we have $\Omega_{10}^\prime=0.045\Omega_{20}$ and hence
$V_y=4.9E_R$ (its slight change due to the variation of $\Omega_{10}^\prime$ is negligible).
For observation, one can measure the in-situ shift of a cloud of atoms~\cite{Nakajima2016,Bloch2016,Qian2011,Lei2013PRL}
over a cycle, which takes 75ms if $\omega=0.01$,
much shorter than the life time (several seconds) of the achieved topological gases in the experiment~\cite{Shuai2017}.

In the 3D case, we can apply two independent sets of the setup proposed in Ref.~\cite{Yong2016PRA} for
realization of a Weyl nodal ring. Here, the scheme is optimized by using the linearly polarized laser beams as
shown in Fig.~\ref{Fig3}(c) and (d). In the first set, two pairs of Raman processes are utilized
to generate the off-diagonal optical lattices.
One pair has
the Rabi frequencies: [$\bar{\Omega} _{1}=\bar{\Omega}_{10}\cos
(k_{Ly}r_{y})e^{-ik_{Lz}r_{z}/2}$, $\bar{\Omega} _{2}=-i\bar{\Omega}_{20}\sin
(k_{Lx}r_{x})e^{ik_{Lz}r_{z}/2}$], and the other pair:
[$\bar{\Omega} _{1}^{\prime }=\bar{\Omega} _{10}\cos
(k_{Ly}r_{y})e^{ik_{Lz}r_{z}/2}$, $\bar{\Omega} _{2}^{\prime }=-i\bar{\Omega}
_{20}\sin (k_{Lx}r_{x})e^{-ik_{Lz}r_{z}/2}$].
In the second set, two pairs of Raman laser beams are employed to
engineer the other off-diagonal optical lattices.
The Rabi frequencies for one pair are
[$\tilde{\Omega} _{1}=\tilde{\Omega} _{10}\cos
(k_{Ly}r_{y})e^{-ik_{Lz}r_{z}/2}$, $\tilde{\Omega} _{2}=i\bar{\Omega}
_{10}\cos (k_{Lx}r_{x})e^{ik_{Lz}r_{z}/2}$], while for the other pair
[$\tilde{\Omega} _{1}^{\prime }=\tilde{\Omega} _{10}\cos
(k_{Ly}r_{y})e^{ik_{Lz}r_{z}/2}$, $\tilde{\Omega} _{2}^{\prime }=i\bar{\Omega}
_{10}\cos (k_{Lx}r_{x})e^{-ik_{Lz}r_{z}/2}$].
In an experiment, by taking $\bar{\Omega}_{10}=2\pi\times0.14\text{GHz}$ and $\bar{\Omega}_{20}=\bar{\Omega}_{10}/4$,
we have $V_x\approx V_y\approx 3.2E_R$. Another laser beam is required to create
an optical lattice along $z$ with $V_z=3.2E_R$.
Using the geometry of lasers with $k_{Lx}=k_{Ly}=k_{Lz}=\sqrt{4/5}k_R$, we have $J_x=0.07E_R$ and $h_y=-0.57V_{Zy0}$.
Similar to the 2D scenario, the dynamical 4D Weyl nodal ring with the topological pumping can be
implemented by tuning $h_y$ and $h_z$ with a period being $86\text{ms}$ if $\omega=0.01$.

{\LARGE \textbf{Discussion}}

Dynamical Weyl points and 4D Weyl nodal rings may also be implemented in solid-state materials
by applying circularly polarized lights to 2D Dirac materials or 3D nodal line semimetals, respectively.
It allows us to engineer an effective time-independent Hamiltonian obtained by removing the fast
oscillating terms, when the frequency of lights is much larger than other energy scales;
this method has been proposed to generate Weyl points from Weyl nodal line semimetals~\cite{WangPRL2016}.
Additionally, slowly varying the light intensity allows us to control the Hamiltonian for observation
of the topological pump. Despite
the possibility, we have to say that a very exquisite protocol is required for realization of
dynamical Weyl points and dynamical Weyl nodal rings in solids.

Versatile controllability of cold atoms also manifest in tuning the short-range interactions by Feshbach resonances,
which can be tuned to zero. For weak interactions, a mean-field estimate suggests
that the interaction may induce a $\sigma_z$ term~\cite{Yong2016typeii}, thereby shifting the locations of
dynamical Weyl points and 4D Weyl nodal lines and changing the amount of pumped particles.
For strong interactions, previous study suggests that a Weyl point may become Mott gapped
while preserve a gapless collective excitation~\cite{Nagaosa2016}. Whether the value of pumped particles will
be strongly compromised depends on the the density of state
around zero energy, which deserves future exploration.

In summary, we have demonstrated the existence of a dynamical Weyl point and 4D
Weyl nodal ring in 2D and 3D systems, respectively. We find that these systems give
rise to the non-quantized topological pump and the amount of transported
particles can be tuned continuously by controlling experimental parameters.
We finally propose an experimental scheme to realize the dynamical Weyl point
and 4D Weyl nodal ring and to observe their corresponding tunable pump, which
paves the way for their future experimental observation. Our finding opens
a field for studying dynamical gapless phenomena; future direction may
include seeking other dynamical gapless phenomena, such as dynamical
Yang monopoles.

\medskip {\LARGE \textbf{Methods}}

\textbf{Tight-binding Hamiltonian in the momentum space}:
We can write down the Hamiltonian in the second quantization language
\begin{equation}
H_{II}=\int d\mathbf{r}\hat{\psi}^{\dagger }(\mathbf{r})H_{C}\hat{\psi}(\mathbf{r%
}),
\label{HSec}
\end{equation}%
where $%
\hat{\psi}(\mathbf{r})=[%
\begin{array}{cc}
\hat{\psi}_{\uparrow }(\mathbf{r}) & \hat{\psi}_{\downarrow }(\mathbf{r})%
\end{array}%
]^{T}$ with $\hat{\psi}_{\sigma }(\mathbf{r})$ [$\hat{\psi}_{\sigma
}^{\dagger }(\mathbf{r})$] being an annihilation (creation) operator for spin $\sigma $
($\sigma =\uparrow ,\downarrow $), which satisfies the
anti-commutation or commutation relation $[\hat{\psi}_{\sigma }(\mathbf{r}),%
\hat{\psi}_{\sigma ^{\prime }}^{\dagger }(\mathbf{r}^{\prime })]_{\pm
}=\delta _{\sigma \sigma ^{\prime }}\delta (\mathbf{r}-\mathbf{r}^{\prime })$
for fermionic atoms ($+$) or bosonic atoms ($-$), respectively. The field
operator can be approximated by
\begin{equation}
\hat{\psi}_{\sigma }(\mathbf{r})\approx\sum_{{\bf x},\sigma }W_{{\bf x}}\hat{c}%
_{{\bf x},\sigma },
\label{EWn}
\end{equation}%
where $W_{\bf x}$ is the Wannier function for $h_z=V_N=0$ located at the site ${\bf x}=\sum_\nu j_\nu a_\nu {\bf e}_\nu$
with $\nu=x,y$ in 2D ($\nu=x,y,z$ in 3D) for
the lowest band, and $\hat{c}_{{\bf x},\sigma}$ is the operator annihilating a particle with spin $\sigma$ at a site ${\bf x}$.

Substituting Eq.~(\ref{EWn}) into Eq.~(\ref{HSec}) and keeping only the nearest neighbor hopping terms yields the tight-binding
Hamiltonian
\begin{eqnarray}
H_{TB}=&&\sum_{\bf x} \left[-\sum_{\nu}\left(J_{\nu}\hat{c}_{\bf x}^{\dagger}\hat{c}_{{\bf x}+a_\nu {\bf e}_\nu}+H.c.\right)+
h_{z}\hat{c}_{{\bf x}}^{\dagger}\sigma_{z}\hat{c}_{\bf x}\right] \\ \nonumber
&&+\sum_{\bf x}g_{\bf x}\left(-J_{SO}\hat{c}_{\bf x}^{\dagger}\sigma_{y}\hat{c}_{{\bf x}+a_x {\bf e}_x}
+H.c.+h_{y}\hat{c}_{\bf x}^{\dagger}\sigma_{y}\hat{c}_{\bf x}\right),
\end{eqnarray}
where $\hat{c}^\dagger_{\bf x}=(\hat{c}^\dagger_{{\bf x},\uparrow},\hat{c}^\dagger_{{\bf x},\downarrow})$ and $g_{\bf x}=(-1)^{j_{x}+j_{y}}$ in 2D [$g_{\bf x}=(-1)^{j_{x}+j_{y}+j_z}$ in 3D].
For more details, we refer to Ref.~\cite{Yong2016PRA,Yong2016typeii} for derivation of the model and verification for its validity.
Using the transformation $\hat{a}_{{\bf x} \uparrow}=g_{\bf x}\hat{c}_{{\bf x}\uparrow}$ and
$\hat{a}_{{\bf x}\downarrow}=\hat{c}_{{\bf x}\downarrow}$, we recast the model to the form
\begin{eqnarray}
H_{TB}=&&\sum_{\bf x}[(\sum_{\nu}J_{\nu}\hat{a}_{\bf x}^{\dagger}\sigma_{z}\hat{a}_{{\bf x}+a_\nu{\bf e}_\nu}+
iJ_{SO}\hat{a}_{\bf x}^{\dagger}\sigma_{x}\hat{a}_{{\bf x}+a_x{\bf e}_x} \nonumber \\
&&+H.c. )
+h_{z}\hat{a}_{\bf x}^{\dagger}\sigma_{z}\hat{a}_{\bf x}+h_{y}\hat{a}_{\bf x}^{\dagger}\sigma_{y}\hat{a}_{\bf x}].
\end{eqnarray}
This Hamiltonian can be written in the momentum space as $H_{TB}=\sum_{\bf k}\hat{a}_{\bf k}^\dagger H({\bf k})\hat{a}_{\bf k}$,
where $H({\bf k})$ is given in Eq.~(7).

\textbf{Acknowledgements}: We thank F. Mei for helpful discussions.
This work was supported by the start-up program of Tsinghua University (533303002) and
the Ministry of Education and the National key Research
and Development Program of China (2016YFA0301902).

\textbf{Author contributions} Y. Xu conceived the idea and supervised the project.
Y.-B. Yang and Y. Xu obtained the numerical results.
All authors took part in discussion of the results and experimental scheme.
Y. Xu and L.-M. Duan wrote the manuscript.

\textbf{Competing financial interests}

The authors declare no competing financial interests.

\end{document}